\documentclass[conference]{IEEEtran}
\IEEEoverridecommandlockouts
\usepackage{cite, float}
\usepackage{amsmath,amssymb,amsfonts}
\usepackage{algpseudocode}
\usepackage{algorithmicx}
\usepackage{graphicx}
\usepackage{textcomp}
\usepackage{xcolor}
\usepackage[caption=false]{subfig}
\usepackage{titlesec}
\def\BibTeX{{\rm B\kern-.05em{\sc i\kern-.025em b}\kern-.08em
    T\kern-.1667em\lower.7ex\hbox{E}\kern-.125emX}}
\usepackage{ifpdf}
\ifpdf
\usepackage{hyperref}
\fi
\titlespacing*{\section}
{0pt} % Left margin
{1.5em plus 2pt minus 2pt} % Above, with adjustment range
{1em plus 2pt minus 2pt} % Below

\titleformat{\section}
{\centering\large\bfseries} % Format: large font size, bold, centering
{\thesectiondis} % Section number
{0.5em} % Space between number and title
{}

\titlespacing*{\subsection}
{0pt} % Left margin
{1em plus 2pt minus 2pt} % Above
{0.5em plus 2pt minus 2pt} % Below

\titleformat{\subsection}
{\bfseries} % Format:  bold
{\thesubsectiondis} % Subsection number
{0.5em} % Space between number and title
{}

\begin{document}

\title{Collaborative Access Control for IoT - A Blockchain Approach}

\author{\IEEEauthorblockN{Huang, Yongtao}
\IEEEauthorblockA{\textit{University of Texas at Dallas} \\
huang.yongtao@utdallas.edu}
\and
\IEEEauthorblockN{I-Ling Yen}
\IEEEauthorblockA{\textit{University of Texas at Dallas} \\
ilyen@utdallas.edu}
\and
\IEEEauthorblockN{Farokh Bastani}
\IEEEauthorblockA{\textit{University of Texas at Dallas} \\
bastani@utdallas.edu}
}

\maketitle
\thispagestyle{plain}
\pagestyle{plain}

\begin{abstract}
The Internet of Things (IoT) necessitates robust access control mechanisms to secure a vast array of interconnected devices. Most of the existing IoT systems in practice use centralized solutions. We identify the problems in such solutions and adopt the blockchain based decentralized access control approach. Though there are works in the literature that use blockchain for access control, there are some gaps in these works. We develop a blockchain embedded access control (BEAC) framework to bridge the gaps. 
First, blockchain based solutions for access control require an enabling P2P network while existing P2P overlays do not support some required features. We develop a novel P2P infrastructure to seamlessly support our BEAC framework. Second, most of the works consider blockchain based access control for a single access control model, and we develop a generic blockchain mechanism and show that it can support the embedding of various access control models. Finally, existing works adopt existing blockchain mechanisms which may incur a high communication overhead. We develop a shortcut approach to improve the number of message rounds in the access protocol. Our experiments demonstrate the efficacy of our system, showing that the shortcut mechanism can reduces access time by approximately 43\%. 
\end{abstract}

%\begin{IEEEkeywords}

%\end{IEEEkeywords}

\section{Introduction}
Internet of Things (IoT) has become indispensable in connecting a diverse range of devices from wearable, home, industrial, surveillance, agricultural, and smart city sensors to mobile devices, automobile cameras, drones, etc., which significantly enhances convenience and efficiency of daily operations across numerous sectors. However, this rapid expansion introduces challenges, particularly in security and access control, as many IoT devices handle sensitive information. 

In practical deployments of IoT systems, access control usually relies on simple encryption-based handshakes for owner access or Single Sign-On (SSO) gateways for user authentication. Handshakes are straightforward but insufficient for complex sharing, whereas the SSO approach requires a stable internet connection, causing issues during outages - a significant concern in smart homes where centralized authentication can block accesses during internet disruptions, even when users and devices are on the same LAN.
While advanced users might overcome these challenges with dynamic DNS, reverse proxies, and HTTPS certificates, these solutions can be too complex and potentially insecure for average users. 
Some other manufacturers streamline device management through proprietary social networks or apps, which mandates account registration and family-joins and, subsequently, poses security and privacy risks.

Centralized or vendor-specific approaches for access control in IoT systems often result in inflexibility and limitations. Considering IoT's inherently decentralized structure, there's a growing trend in academic literature for peer-to-peer (P2P) access control with blockchain mechanisms.
In \cite{chen_blockchain_2017}, a Bitcoin-based approach is introduced for enforcing Attribute-based Access Control (ABAC) via transactions. This method supports basic policy management tasks such as creation, update, and revocation, and uniquely enables the transfer of access rights through Right Transfer Transactions (RTT). 
In \cite{novo_blockchain_2018} \cite{alphand_iotchain_2018} \cite{zhang_smart_2019}, more advanced blockchain based access control frameworks on Ethereum have been developed, which employ smart contracts and can be applied to multiple access control models.
However, though utilizing blockchain to oversee policies and access transactions has been investigated, there are several gaps in the literature of this direction toward fully realizing such approaches.

Firstly, the literature in blockchain embedding for access control does not consider the P2P network design to effectively support the proposed mechanisms. In fact, IoT devices behind dynamic IPs and firewalls, especially when accessed from similar networks, present significant connectivity challenges. Existing P2P solutions like Kademlia\cite{goos_kademlia_2002}, BitTorrent, and Gnutella\cite{bordignon_gnutella_2001} provide partial remedies but are insufficient for IoT's unique requirements. A common limitation within these P2P networks is the inability for two peers, each situated in disparate, firewalled subnets, to forge a direct connection.
Moreover, the use of dynamic IP, network tunneling, and port forwarding, etc., as commonly used by many IoT devices hidden in localized subdomains, results in a single peer being bound to multiple IP addresses, making many P2P systems that use IP addresses to define peer identities infeasible.   

The second gap in the blockchain based access control literature is in handling multiple access control models. Most of the existing works consider embedding access control in blockchain for a single access control model. The mechanism is insufficient for many large-scale IoT systems where different access control models are used in different IoT domains. 

A third issue is with the performance concerns. Most of the related works directly use an existing blockchain mechanism without considering the communication overhead.  While blockchain integration offers advancements in access control models and smart contract deployment, the required number of messaging rounds in the blockchain protocols can lead to significant latency. 

\begin{figure}[htbp]
	\centering\noindent\includegraphics[width=0.38\textwidth]{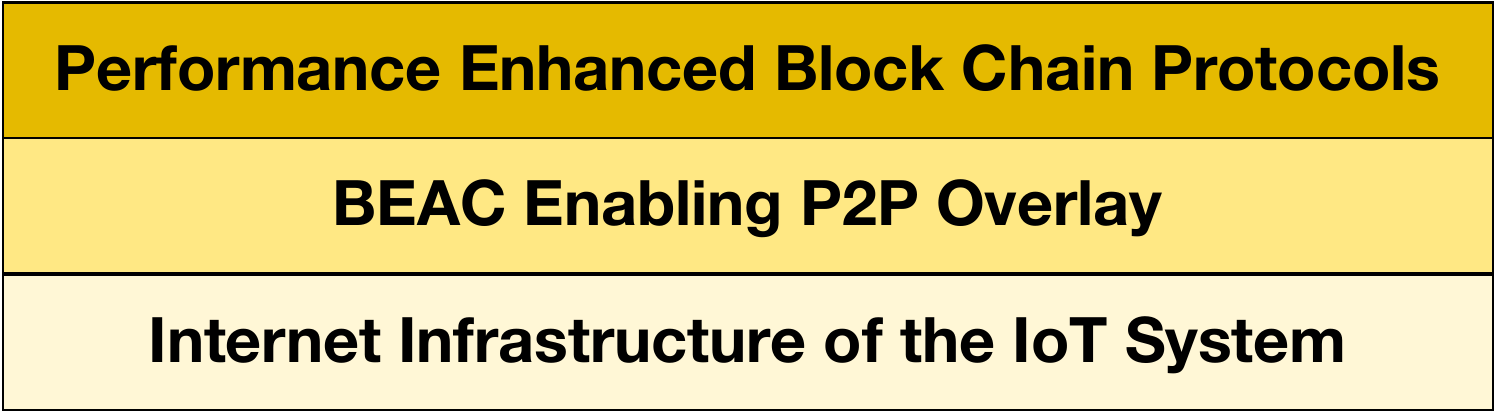}
	\caption{3-layer Framework}
	\label{3_layer}
\end{figure}

To address the gaps discussed above, we introduce a novel P2P blockchain-embedded access control (P2P-BEAC) framework designed for decentralized access control in large-scale P2P IoT systems with substantial IoT resource sharing. P2P-BEAC consists of three layers as shown in  \ref{3_layer}.

The bottom layer lays the existing Internet infrastructure, including various nodes that are relevant to the system, such as IoT devices, edge nodes, users, as well as servers that can provide blockchain embedded access control services. 

In the second layer, we leverage libp2p to build an effective P2P infrastructure to enable a single, virtually universal domain which allows end-users to access IoT devices seamlessly across internet and local networks. The system offers advanced networking techniques to break the subdomain and firewall barriers, the IoT service discovery mechanisms to support the  sharing of IoT devices with multiple stakeholders and elevates transparency beyond traditional solutions, and the publish-subscribe protocol to support reliable communication among the system entities for handling accesses and performing access control and validations. 

The third layer is the core blockchain embedded access control (BEAC) unit. It consists of entities that provide the access control services and the corresponding protocols, such as bootstraping, access request handling, and access policy validation protocols. 
Interoperability is one of the focuses in this layer. We design a common blockchain mechanism and demonstrate that it can support multiple access control models and facilitate the implementation of various cross-domain access techniques.
As with other blockchain based access control solutions, this layer also provides a robust crash recovery mechanism that is specific to our protocol, making it versatile for use in many large-scale IoT systems with IoT resource sharing.

The BEAC layer also provides mechanisms for performance improvement to reduce the overhead introduced in blockchain protocols. For access control, we require that both the device (or domain) and the blockchain service to validate the access rights of the request against the policies. With a little additional overhead due to duplicated processing, we enable a significant reduction in the number of message rounds via a shortcut mechanism, which allows full access validation being done in parallel with local access authorization. This can benefit accesses by permanent users, such as the owner, and by recurrent accessors, which occur frequently in real-world access patterns.

In summary, our P2P-BEAC offers enabling P2P techniques to make the realization of blockchain embedded access control feasible in practice, supports multiple access control models, and significantly enhances the performance for blockchain based access control solutions. The design of our P2P-BEAC framework is discussed in the subsequent four sections, with the bottom layer in Section \ref{sec:arc}, the middle layer in Section \ref{sec:p2p}, the core of the BEAC layer in Section \ref{sec:beac}. In Section \ref{sec:perf}, we discuss the shortcut protocol for efficient access validation and analyze its performance. Finally, Section \ref{sec:conclusion} concludes the paper.

\section{Infrastructure Foundations}\label{sec:arc}
\begin{figure}[htbp]
	\centering\noindent\includegraphics[width=0.48\textwidth]{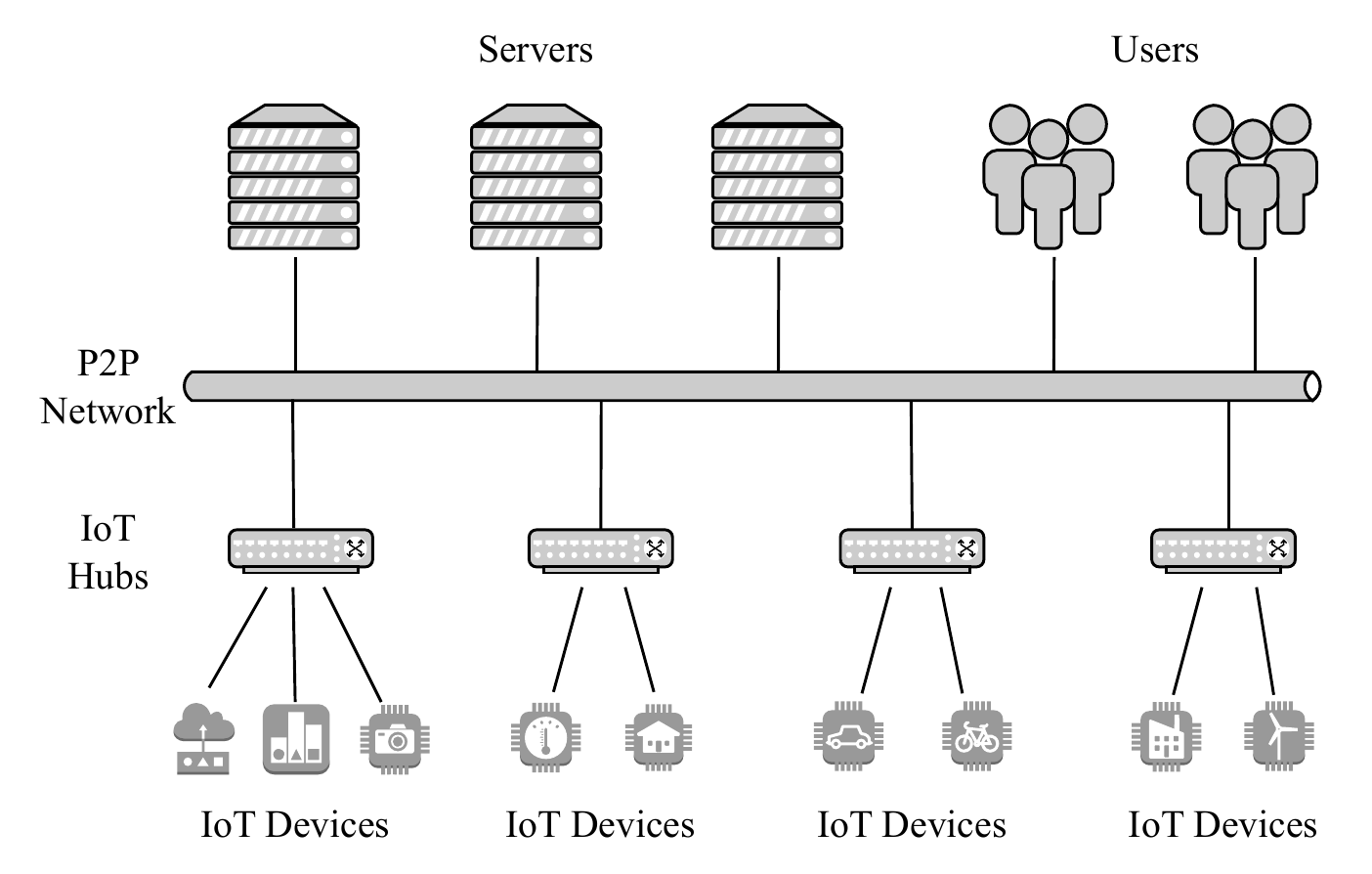}
	\caption{System Architecture}
	\label{fig_overall}
\end{figure}

The system entities that comprise the bottom layer of our P2P-BEAC framework are depicted in Figure \ref{fig_overall} and each entity is further defined and elaborated in the following. 

\textbf{IoT Devices} offer a wide spectrum of IoT services. As how they are placed in the current Internet, we consider them hidden in the subnets behind firewalls.
% from basic sensor data collection, like video streaming or temperature monitoring, to executing advanced interactive tasks such as object recognition and controlling multiple actuators. 
Services provided by the IoT devices are defined via manufacturer-provided templates, which can be used for semantic based service discovery.
Each IoT device can be the basic element for access control and policies can be assignable on a per-service basis.

\textbf{Users} can access the IoT devices in the network according to the access control polices for the devices. A user can represent a diverse range of entities, including an independent individual, an individual in a domain, a software program or a service, or even some IoT entity or its service. 

\textbf{Owners.} Device owner can be an individual or an organization who owns the device and has full access rights to the device. The access rights cannot be revoked except when ownership is transferred. 

\textbf{Domain} is a higher-level entity in the system (e.g., homes, factories, companies), may consist of IoT hubs and IoT devices. IoT devices and their owners shall be in the same domain. Instead of defining policies for individual IoT devices, a domain can host a generic policy for the devices or device groups in the domain. A single IoT device with sufficient capabilities can also choose to form its own domain. 

\textbf{Server Nodes} are relatively powerful computers with high storage capacity or a cluster of computers in a local area network or in the cloud. We make use of these nodes to perform blockchain based access control and maintain blockchain and other system state data. In P2P-BEAC, server nodes serve as blockchain validators and access control hubs.

\section{BEAC-Enabling Peer-to-Peer Overlay}\label{sec:p2p}
%As depicted in Fig.~\ref{fig_overall}, our IoT system is fundamentally anchored in a peer-to-peer (P2P) network. 

\begin{figure}[htbp]
	\centering\noindent\includegraphics[width=0.48\textwidth]{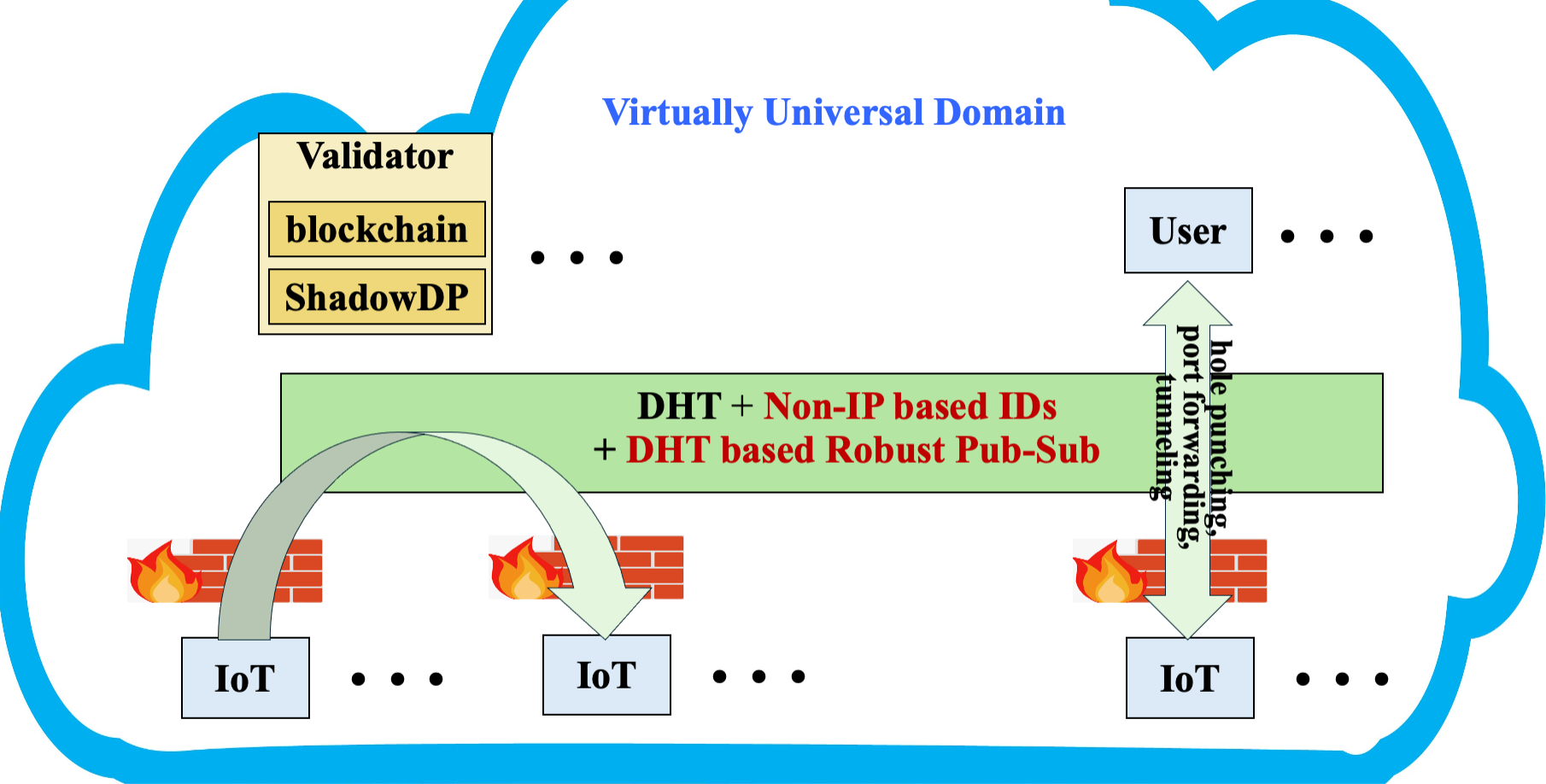}
	\caption{P2P Overlay Network}
	\label{fig:domain}
\end{figure}

% should it be p2p or P2P, make it consistent
Though the design choice of the overlay infrastructure is pivotal for enabling a seamless blockchain embedded access control in a large-scale, resource sharing IoT system, a specific design of the P2P network has not been considered in the literature. 
To fully realize controlled sharing of IoT devices and enforce the associated access control policies, the network must adhere to several critical criteria. 
We leverage the libp2p library from Filecoin\cite{noauthor_libp2pgo-libp2p_2024}
%\cite{psaras_interplanetary_2020} 
to design our P2P overaly network infrastructure to meet these criteria.
The specific criteria and the corresponding solutions are depicted in Figure \ref{fig:domain} and discussed in the subsequent paragraphs.

\subsection{Virtually Universal Domain}
To ensure device accessibility across both internet and intranet environments, the P2P network should remove the barrier of individual sub-domains protected behind firewalls, linking entities into a universal domain to facilitate boundaryless accesses.

We explore multiple P2P implementations and identify their problems in supporting such a virtually universal domain. Finally, our P2P-BEAC framework chooses to leverage some capabilities in libp2p and build upon them to address the challenge. First, our P2P network employs hole punching and fallback relays\cite{seemann_decentralized_2022} in libp2p to ensure that the devices behind firewalls can be accessed by users in some different subnets protected by different firewalls.
We also incorporate port forwarding and virtual IP capabilities via TAP/TUN devices. In conjunction with libp2p's stream multiplexing, this approach enables a peer to access remote IoT device services as though they were on the same local area network (LAN).

%	Additional features such as network tunneling and service forwarding are added to libp2p to accommodate various IoT services by enabling simultaneous service streams over a single transport, drawing inspiration from SPDY\cite{belshe_spdy_2012} (the precursor to HTTP/2\cite{belshe_hypertext_2015}). It seamlessly integrates remote and local device services through network tunnels, achieved through protocol encapsulation within multiplexed streams and port forwarding, ensuring user access is as straightforward remotely as it is locally. To prevent port collisions when accessing multiple services of the same type, such as monitoring array of surveillance cameras, the system utilizes virtual IP addresses provided by a local TUN/TAP device.

\subsection{System Entity Identification and Authentication.} 
\textbf{DHT.} The P2P network is organized by Distributed Hash Tables (DHT) to maintain the mapping between peer identities and associate network addresses.
Rendezvous nodes and Zero Configuration Networking
(zeroconf) are used to boost initial mapping

\textbf{Problem with IP based identity.} Each peer (an IoT device or a user) within the network should be uniquely identifiable. Traditional P2P networks that use IP for creating P2P identities face challenges because multiple IP addresses may be bound to a single peer. 
For example, a peer may have a local IP in the intranet (subdomain hidden from the public network). When communicating with external domains, a public IP address is temporarily mapped to the peer via network address translation (NAT) or port forwarding protocols. For many subdomains, this mapped public IP address may be dynamic, and, hence, each peer may be associated with many IP addresses and each IP address may also be mapped to different peers at different times. Thus, it is infeasible to use IP address as the unique identifier in a P2P system.  

\textbf{Public key for Entity Identification and Authentication.} Our system leverages Elliptic Curve Digital Signature Algorithm (ECDSA) key pairs to define the identify of each peer. The ECDSA based identity, along with pertinent network addresses, is stored in the DHT to enable device discovery. ECDSA also facilitates the ready authentication protocols and the establishments of secure connections among peers. 

%To manage dynamic network topologies, each peer maintains a transient peer store, refreshed during interactions via publish-subscribe protocols, DHT, or direct P2P connections. This store dynamically maps IDs to IP addresses of known peers, facilitating ongoing communication despite network changes.

\subsection{Pub-Sub Communication Protocol.} 
Given dynamic and decentralized nature of P2P networks, a communication protocol is required to support robust message delivery while limiting flooding. The pub-sub  protocol is adopted to support the publish-subscribe communication pattern. The pub-sub communication is  but without a centralized broker or bus.
Also, in access control and blockchain protocols, broadcasting and multi-casting may frequently be needed for data dissemination and event notification. The pub-sub protocol offers reliable delivery of the message to multiple designated destinations\cite{vyzovitis_gossipsub_2020}.

\section{Blockchain Embedded Access Control (BEAC)}\label{sec:beac}
Our blockchain architecture adheres to established standards. For consensus, we employ a state-of-the-art Byzantine Fault Tolerance (BFT) protocol, specifically a 2-chain HotStuff\cite{yin_hotstuff_2019-1} variant known as Ditto\cite{gelashvili_jolteon_2022}. The block data structure is modeled after the JellyFish Merkle tree\cite{gao_jellyfish_2021}, a design borrowed from the Diem\cite{amsden_libra_2020} project.

A key feature of our system's design is its resilience, particularly in maintaining operational integrity despite node failures, i.e., IoT devices and domains can reconstruct their access control states after failures. 
Thus, we choose to embed the access control policies in the blockchain transaction protocol with specific registration requirements.
Our design ensures the seamless addition or removal of hubs as required and offers robust recovery mechanisms from catastrophic device failures, 

Also, engineered for adaptability, our system supports a multitude of domains, each governed by its distinct access control model. We provide demonstrative implementations of Discretionary Access Control (DAC), Attribute-Based Access Control (ABAC), and Role-Based Access Control (RBAC). 

In the next subsection, we first outline the key system entities in the BEAC layer and their roles in implementing BEAC protocols (in Subsection \ref{ssec:entity}).
To illustrate the compatibility of our P2P-BEAC framework, the integration of BEAC with DAC, ABAC, and RBAC are detailed in Subsections \ref{ssec:dac}, \ref{ssec:rbac} and \ref{ssec:abac}). Finally, we further explore the dynamic capabilities offered by smart contracts for a more adaptive access control environment \ref{ssec:dynamicAC} .

\subsection{BEAC Preliminary}\label{ssec:entity}
Besides the entities discussed in Section \ref{sec:arc}, some additional entities for the BEAC layer are introduced below.

\begin{itemize}
\item\textbf{Validator Cluster:} At the core of our system is a consortium blockchain, a design choice that balances chain update latency against fairness. Validators, which are dedicate servers, maintain the complete state of the blockchain and its storage. As their designation implies, these entities are tasked with appending new blocks to the blockchain, utilizing Byzantine Fault Tolerance (BFT) consensus protocols to ensure network agreement and integrity.

\item\textbf{IoT Hubs:} Serving as control nodes within specific domains (e.g., homes, factories, companies), IoT hubs preserve domain local state of the blockchain, enabling them to partially verify the chain's integrity. While they can propose new blocks to the validators, only validators have the authority to determine the acceptance of these proposals. Beyond their blockchain role, IoT hubs act as bootstrapping and rendezvous points for IoT devices within their domain, facilitating device network integration and communication.
\end{itemize}

Since we embed access control policies in the blockchain, it is necessary to bootstrap the IoT devices and domains. The concept of ownership and the registration requirements are discussed below.

\begin{itemize}
\item\textbf{Device Registration:} Any device is required to register itself in order to join the network. The registration request includes the device's and its owner's signed fingerprints. Once registered, the device record in blockchain is immutable. Changes in ownership require releasing by the current owner and re-registration by the new owner. The newly registered device will have different fingerprint. 
	
\item\textbf{Ownership:} Device owner has full access. The access cannot be revoked except when ownership is transferred. For Mandatory Access Control (MAC), device ownership rights are streamlined to a selected group of administrative users, primarily for failure recovery. 	Inspired by Linux Security Modules framework\cite{wright_linux_2002}, our system supports various access control models across different domains.
	
\item\textbf{Domain Registration:} Domains, their rules, and owners must register immutably. Global domains can set ownership to a generic address, analog to nobody user in Linux, allowing open access while preventing spam by restricting registration to consortium members.
\end{itemize}

\subsection{Discretionary Access Control (DAC)}\label{ssec:dac}
Beside default domain and device registration processes, here are default access permission types associated with a device:
\begin{itemize}
	\item\textit{LIST.} Permission to view device details, including its manufacture information and services provided.
	\item\textit{CHMOD.} Permission to grant or deny access to other user or device. Special user nobody and everybody can be used.
	\item\textit{EXECUTE.} Permission to execute a device service, or create a tunnel to a running one.
\end{itemize}

Permissions in our system are managed through Access Control Lists (ACL) located either on a device hub or directly on devices with hub capabilities. The blockchain records permission transactions rather than the ACL itself. Execution of a permission request by the device or hub generates a token for the user, which is then authenticated and ratified by validators before being logged on the blockchain. These tokens can be configured for single use, have an expiration time, or be set as permanent such as for device owners, streamlining access for recurrent visits. Concurrently, a session key, paired with the token, is dispatched to the device providing the requested service, which then grants access upon matching the session key and token.

Drawing inspiration from Kerberos\cite{neuman_kerberos_1994}, our model leverages blockchain for token distribution, effectively addressing the single-point-of-failure issue inherent in Kerberos's ticket-granting service.

Storing the ACL on hub or device is practical, since offline devices negate the need for access. Yet, in an enterprise context, this centralization could introduce a single point of failure. To mitigate this, ACLs can be replicated across multiple hubs within the same enterprise domain and synchronized with the blockchain, enhancing reliability. The replication and load-balancing policy for ACL is beyond scope of this paper.

The table outlines the types of transactions recorded on the blockchain for basic Discretionary Access Control (DAC), forming a comprehensive transaction log:
\begin{table}[htbp]
	\caption{DAC Transactions}
	\begin{center}
		\begin{tabular}{|c|c|c|c|}
			\hline
			\multicolumn{4}{|c|}{\textbf{Domain Registration}} \\
			\hline
			issuer & domain  & owner  & access control contract \\
			\hline
			\multicolumn{4}{|c|}{\textbf{Device Registration}} \\
			\hline
			issuer & device  & owner  & services \\
			\hline
			\multicolumn{4}{|c|}{\textbf{Device Revocation}} \\
			\hline
			issuer & device  & \multicolumn{2}{|c|}{} \\
			\hline
			\multicolumn{4}{|c|}{\textbf{Permission Granted}} \\
			\hline
			issuer & user  & device/service  & permission type  \\
			\hline
			\multicolumn{4}{|c|}{\textbf{Permission Revoked}} \\
			\hline
			issuer & user  & device/service  &permission type \\
			\hline
		\end{tabular}
		\label{tab_dac}
	\end{center}
\end{table}

Each record is signed by issuer's private key with timestamp and checksum, therefore cannot be tampered.

For simplicity, the pseudo code given below for ACL reconstruction assumes that the blockchain records are consistent and omits detailed validation steps like issuer fingerprint verification, record uniqueness checks, and permission conflict resolution, etc. Also, to generalize, the code below considers a device tree instead of a single device. Specifically, function $BuildDeviceTree$ reconstructs the local device tree, providing a consistent basis for all subsequent device management operations within the network.

\begin{algorithmic}[1]
	\State Initial State:
	\Statex $device\_tree := \emptyset$
	\Statex $known\_id\_map := \{\}$	\Comment{id to device tree node mapping}
	\Statex $activation\_user$	\Comment{Require hub owner to bootstrap}
	\Statex
	\Function{BuildDeviceTree}{record}
	\State $genesis\_found := false$ \Comment{There is only one genesis record for consistent records}
	\Repeat
	\If{$record$ is DomainRegistration}
	\State $genesis\_found$ 
	\Statex $= BootstrapDomain(record)$ 
	\EndIf
	\State $record = record.next$
	\Until{$genesis\_found == true$}
	\Loop
	\If{$record$ is DeviceRegistration}
	\State $HandleDeviceAdd(record)$
	\ElsIf{$record$ is DeviceRevocation}
	\State $HandleDeviceDel(record)$
	\ElsIf{$record$ is PermissionGranted}
	\State $HandlePermAdd(record)$
	\ElsIf{$record$ is PermissionRevoked}
	\State $HandlePermDel(record)$
	\EndIf
	\State $record = record.next$
	\EndLoop
	\EndFunction
	\Statex
	
	\Procedure{BootstrapDomain}{record}
	\If{$record.owner == activation\_user$}
	\State $device\_tree.root = record.domain$
	\State\Return true
	\EndIf
	\State\Return false
	\EndProcedure  
	\Statex
	
	\Procedure{HandleDeviceAdd}{record}
	\If{$record.owner \in known\_id\_map$}
	\State $parent := known\_id\_map[record.owner]$
	\State insert $record.device$ under  $parent$
	\Statex\Comment{Device is always inserted as leaf for consistent records}
	\EndIf
	\EndProcedure
	\Statex
	
	\Procedure{HandleDeviceDel}{record}
	\If{$record.device \in known\_id\_map$}
	\State $parent := known\_id\_map[record.device.owner]$
	\State delete $record.device$ from  $parent$
	\Statex\Comment{Device is always deleted from leaf for consistent records}
	\EndIf
	\EndProcedure
	\Statex
	
	\Procedure{HandlePermAdd}{record}
	\If{$record.device \in known\_id\_map$}
	\State $node := known\_id\_map[record.device]$
	\If{$record.service$}	\Comment{Permission is service level}
	\State add $(record.user, record.type)$ 
	\Statex to  $node.services[record.service].acl$
	\Else	\Comment{Permission is device level}
	\State add $(record.user, record.type)$ to $node.acl$
	\EndIf
	\EndIf
	\EndProcedure
	\Statex
	
	\Procedure{HandlePermDel}{record}
	\If{$record.device \in known\_id\_map$}
	\State $node := known\_id\_map[record.device]$
	\If{$record.service$}	\Comment{Permission is service level}
	\State remove $(record.user, record.type)$ 
	\Statex from  $node.services[record.service].acl$
	\Else	\Comment{Permission is device level}
	\State remove $(record.user, record.type)$
	\Statex from $node.acl$
	\EndIf
	\EndIf
	\EndProcedure
\end{algorithmic}

\subsection{Attribute-Based Access Control (ABAC)}\label{ssec:abac}
ABAC enhances basic DAC by adding a flexible attribute list and context-based access policies. Unlike ACL entries, attributes do not have direct P2P network addresses, necessitating a meta list stored on the IoT hub. This list catalogs all domain-specific attributes and their associations with users. In additional to ACL, static ABAC policies are formulated as tuples of user attribute, device attribute and permission type.

Attributes carry unique IDs for immutability. Removal of an attribute nullifies its ID, ensuring it cannot be reused. New attributes with the same name receive new IDs, maintaining attribute integrity.

When ABAC is used alongside DAC, denials based on attribute lists take precedence over ACL permissions, except in cases involving the device owner.

For an ABAC domain, the transaction records are extended by the following entries:
\begin{table}[htbp]
	\caption{ABAC Transactions}
	\begin{center}
		\begin{tabular}{|c|c|c|c|}
			\hline
			\multicolumn{4}{|c|}{\textbf{New Attribute}} \\
			\hline
			issuer & domain  & attribute & UID \\
			\hline
			\multicolumn{4}{|c|}{\textbf{Delete Attribute}} \\
			\hline
			issuer &  \multicolumn{3}{|c|}{ domain/UID} \\
			\hline
			\multicolumn{4}{|c|}{\textbf{Assign Attribute Device}} \\
			\hline
			issuer &  domain/UID  & \multicolumn{2}{|c|}{device/service} \\
			\hline
			\multicolumn{4}{|c|}{\textbf{Remove Attribute Device}} \\
			\hline
			issuer &  domain/UID  & \multicolumn{2}{|c|}{device/service} \\
			\hline
			\multicolumn{4}{|c|}{\textbf{Assign Attribute User}} \\
			\hline
			issuer &  domain/UID  & \multicolumn{2}{|c|}{user} \\
			\hline
			\multicolumn{4}{|c|}{\textbf{Remove Attribute User}} \\
			\hline
			issuer &  domain/UID  & \multicolumn{2}{|c|}{user} \\
			\hline
			\multicolumn{4}{|c|}{\textbf{Assign Attribute Permission}} \\
			\hline
			issuer & domain/UID(user)   & domain/UID(device) & permission type\\
			\hline
			\multicolumn{4}{|c|}{\textbf{Revoke Attribute Permission}} \\
			\hline
			issuer &  domain/UID(user)   & domain/UID(device) & permission type \\
			\hline
		\end{tabular}
		\label{tab_abac}
	\end{center}
\end{table}

In addition to DAC domains' device tree, ABAC domains maintain an attribute permission list and user-device attribute relations, allowing for attribute-based access control. Similar to DAC, ABAC domains reconstruct from blockchain records, using a straightforward process.

The attribute name is designed for user convenience and does not impact core operations, which rely solely on UID. Validators verify integrity without needing to know the attribute name, allowing for optional encryption of attribute names by the owner for additional privacy. The encryption method is left to the user's discretion.

While this paper focuses on static ABAC policies, more dynamic policies via domain-level smart contracts are possible but not covered here.

\subsection{Role-Based Access Control (RBAC)}\label{ssec:rbac}
RBAC, similar to ABAC, utilizes a meta list maintained on the IoT hub for organizing user roles within a domain, each identified by a unique role ID and mapped to specific users. In environments where RBAC and DAC coexist, denials based on the role permission list take priority over ACL permissions, with the exception being the device owner's access.

Here are transactions in additional to DAC:
\begin{table}[htbp]
	\caption{RBAC Transactions}
	\begin{center}
		\begin{tabular}{|c|c|c|c|}
			\hline
			\multicolumn{4}{|c|}{\textbf{New Role}} \\
			\hline
			issuer & domain  & role & UID \\
			\hline
			\multicolumn{4}{|c|}{\textbf{Delete Role}} \\
			\hline
			issuer &  \multicolumn{3}{|c|}{ domain/UID} \\
			\hline
			\multicolumn{4}{|c|}{\textbf{Assign Role User}} \\
			\hline
			issuer &  domain/UID  & \multicolumn{2}{|c|}{user} \\
			\hline
			\multicolumn{4}{|c|}{\textbf{Remove Role User}} \\
			\hline
			issuer &  domain/UID  & \multicolumn{2}{|c|}{user} \\
			\hline
			\multicolumn{4}{|c|}{\textbf{Assign Role Permission}} \\
			\hline
			issuer & domain/UID  & device/service & permission type\\
			\hline
			\multicolumn{4}{|c|}{\textbf{Revoke Role Permission}} \\
			\hline
			issuer &  domain/UID   & device/service & permission type \\
			\hline
		\end{tabular}
		\label{tab_rbac}
	\end{center}
\end{table}

Similar to ABAC and DAC, the algorithm is straightforward. A flat RBAC domain can be re-built from blockchain.

For lattice hierarchical role structures, the system uses transactions for role addition and removal. Adjusting a role hierarchy requires temporarily detaching it from base elements, making upward changes to the target role, then reattaching roles. Executed as a batch transaction, this method prevents UID changes during temporary detachments, preserving the role's integrity and allowing for dynamic modifications.
\begin{table}[htbp]
	\caption{Hierarchical RBAC Transactions}
	\begin{center}
		\begin{tabular}{|c|c|c|}
			\hline
			\multicolumn{3}{|c|}{\textbf{Add Role Hierarchy}} \\
			\hline
			issuer & domain/UID (parent)  & domain/UID (child) \\
			\hline
			\multicolumn{3}{|c|}{\textbf{Remove Role Hierarchy}} \\
			\hline
			issuer & domain/UID (parent)  & domain/UID (child) \\
			\hline
		\end{tabular}
		\label{tab_pub_rbac}
	\end{center}
\end{table}

\subsection{Dynamic Access Control}\label{ssec:dynamicAC}
%\textbf{One-Time Access.}
One-time access tokens serve well for temporary access. They're useful when, for example, an owner wants to give a guest brief access to a device without adding them as a permanent user. Utilizing time-based one-time passwords (TOTP)\cite{mraihi_totp_2011}, the device or its hub creates an encrypted token and returns a short-digit passphrase with the owner. The owner then relays this passphrase to the guest through an out-of-band method like an SMS message or verbally. The guest's client decrypts the token with the passphrase, gaining temporary device access. 

TOTP's configurable settings enable customization of both the duration and the number of allowed visits.

%\textbf{Cross-Domain Access Control.}
%In our system, the global uniqueness of user fingerprints simplifies cross-domain access and user role-mapping. This design allows for straightforward identification and authentication of users across different domains, facilitating seamless cross-domain access control.  
%\begin{figure}[htbp]
%\centering\noindent\includegraphics[width=0.30\textwidth]{domain.pdf}
%	\caption{Cross Domain Access}
%	\label{fig_domain}
%\end{figure}

\section{Performance Optimization via Shortcut Access}\label{sec:perf}
For scenarios where access is predictably assured—such as for device owners or permanent users—the IoT hub can expedite this process. It issues the access token directly to the requester while simultaneously initiating background verification. Additionally, if the ACL-holding device is also the service provider, it can immediately facilitate service connection. This shortcut access saves round trip time over internet to validators and time for validators to vote and update blockchain.

The standard full path process for connecting to a device service involves several steps: locating the device's Access Control List (ACL) via the P2P network, requesting access from the ACL holder, verifying the request through the device's hub and validators. Upon successful validation, an access token is issued. Once the token is recorded on the blockchain, it is used to access the desired service.
The procedure for full path access are given as follows.
\begin{enumerate}
	\item\textbf{Connection Initiation:} User's request publication to the target hub incurs a network time of $T_{int}$.

	\item\textbf{Hub Verification:} Verification and token generation by the hub, followed by validator submission, require $T_{int}$
	
	\item\textbf{Validator Consensus:} Time for consensus can be optimal at $5\times T_{int}$\cite{gelashvili_jolteon_2022} or extend to $E(R+4)\times T_{int}$ in worst condition\cite{lu_dumbo-mvba_2020}.
	
	\item\textbf{Token Acquisition:} Retrieving the token from the blockchain takes $T_{int}$.
	
	\item\textbf{Service Initiation:} The pairing session key is sent to device $T_{loc}$. User initiates direct p2p connection to device, starting the service protocol. This operation takes $T_{p2p}$.
\end{enumerate}

\subsection{Shortcut}
For scenarios where access is predictably assured—such as for device owners or permanent users—the IoT hub can expedite this process. It issues the access token directly to the requester while simultaneously initiating background verification. Additionally, if the ACL-holding device is also the service provider, it can immediately facilitate service connection. This shortcut access saves round trip time over internet to validators and time for validators to vote and update blockchain.
\begin{figure}[htbp]
	\centering\noindent%
	\subfloat[Full Path]{\includegraphics[width=0.23\textwidth]{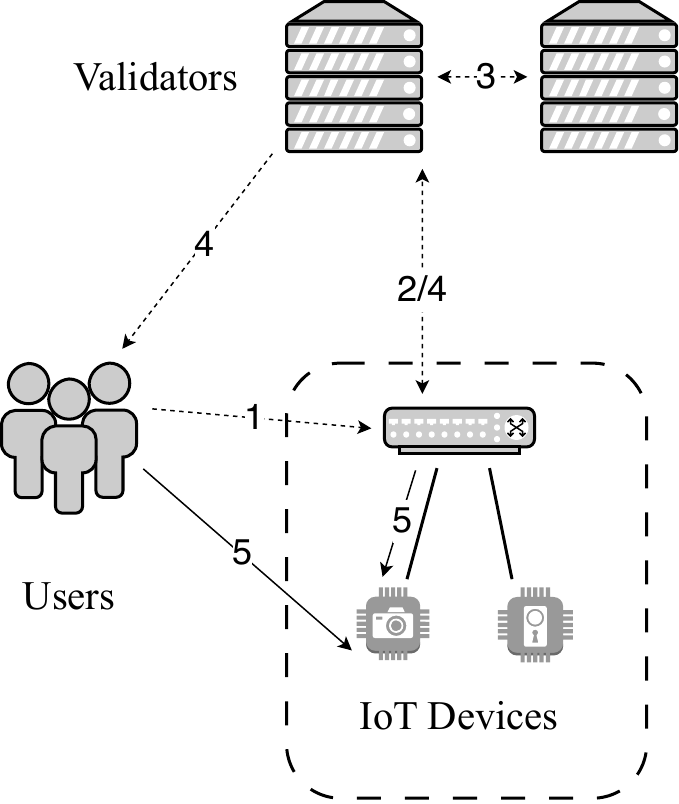}}\hfil%
	\subfloat[Internet Shortcut]{\includegraphics[width=0.23\textwidth]{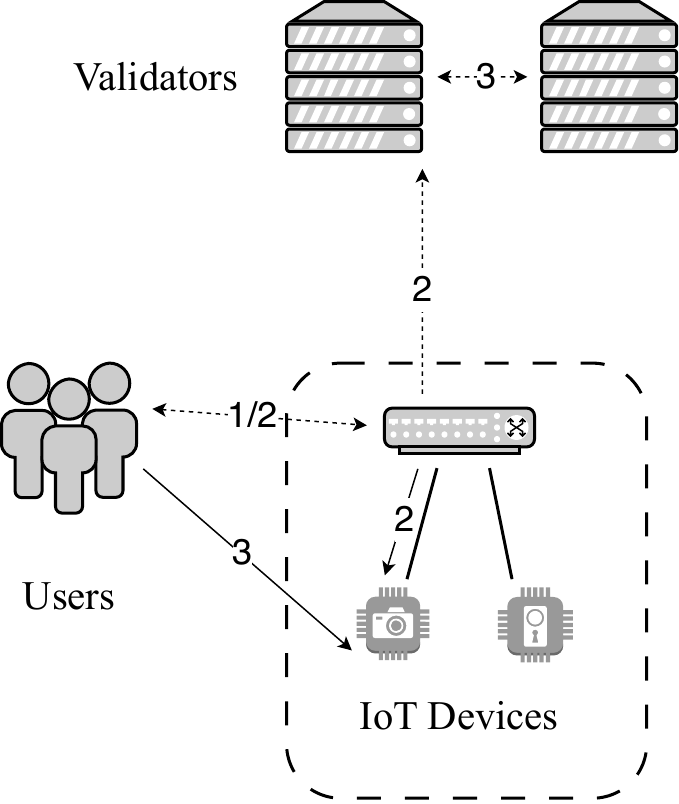}}\\
	\subfloat[Local Shortcut]{\includegraphics[width=0.23\textwidth]{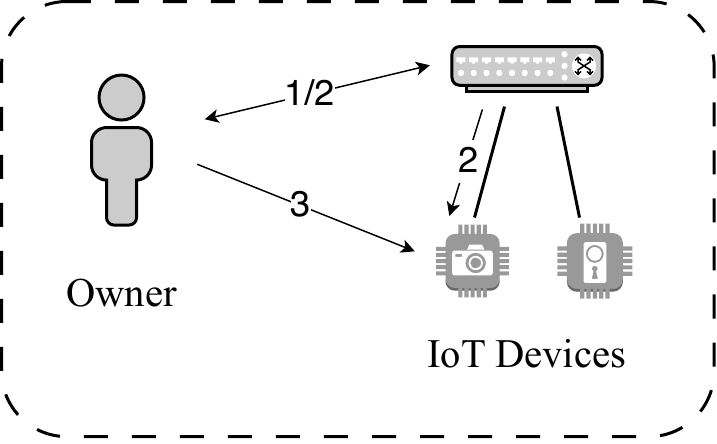}}
	\caption{Full Path and Shortcut Access}
	\label{full_short}
\end{figure}

For shortcut access, we need to consider those that are in the local network and those through the Internet.
The steps for Shortcut access via the Internet are enlisted below. 
\begin{enumerate}
	\item\textbf{Connection Initiation:} Parallel request publication and p2p connection initiation with the device, timed at $T_{int}$. Hole punching may takes longer than retrieving token.
	\item\textbf{Hub Verification:} Immediate token creation and background validation submission by the hub, requiring $T_{int}$, with the session key sent taking $T_{loc}$.
	\item\textbf{Service Initiation:} Service activation using the established p2p channel and the token is at $max(T_{p2p}-2\times T_{int}, 0)$.
\end{enumerate}

Below gives the procedure for shortcut access within the local network. 
\begin{enumerate}
		\item\textbf{Token Retrieval:} Direct connection to the hub for token request takes $T_{loc}$.
		\item\textbf{Hub Verification:} Hub's request verification and token issuance parallel to sending the session key takes $T_{loc}$.
		\item\textbf{Service Initiation:} Device service initiation with the token requires another $T_{loc}$.
\end{enumerate}

To illustrate the full path and shortcut protocols, the above procedures are contrasted in Figure \ref{full_short}

While trivial for static DAC systems, implementing shortcut access in more complex access control models requires careful consideration to maintain system integrity and security.

\subsection{Performance Analysis}
To quantify the time consumed during each operation in the full path access procedure, we consider the internet round trip time through pub-sub ($T_{int}$) and the local network round trip time ($T_{loc}$) through libp2p.

Testing conducted by libp2p developers indicates that pub-sub latency typically remains under 150ms, with the 99th percentile ($P_{99}$) reaching up to 165ms across a network of 1000 nodes using gossipsub-v1.1\cite{vyzovitis_gossipsub-v11_2020}. Local network round trip time usually falls below 1ms.

For initial network connection, bootstrapping is necessary, utilizing methods such as DHT, zeroconf, or a rendezvous point from either a static list or prior peerstore. While the time required for bootstrapping can vary significantly, this process occurs only once in the background and, as such, is not included in our time calculations.

To navigate firewalls for peer-to-peer (p2p) connections, techniques include manual or UPnP port forwarding, hole punching, or relying on a fallback if hole punching is unsuccessful. Port forwarding is immediate, but hole punching requires wo communication stages post a temporary relayed connection\cite{seemann_decentralized_2022}. Tested by libp2p developers, averages connection time is 0.89s ($P_{50}$) and can take up to 7.78s ($P_{99}$) for successful connection establishment ($T_{p2p}$)\cite{ipfs_libp2p_2022}.

\begin{figure}[H]
	\centering\noindent\includegraphics[width=0.48\textwidth]{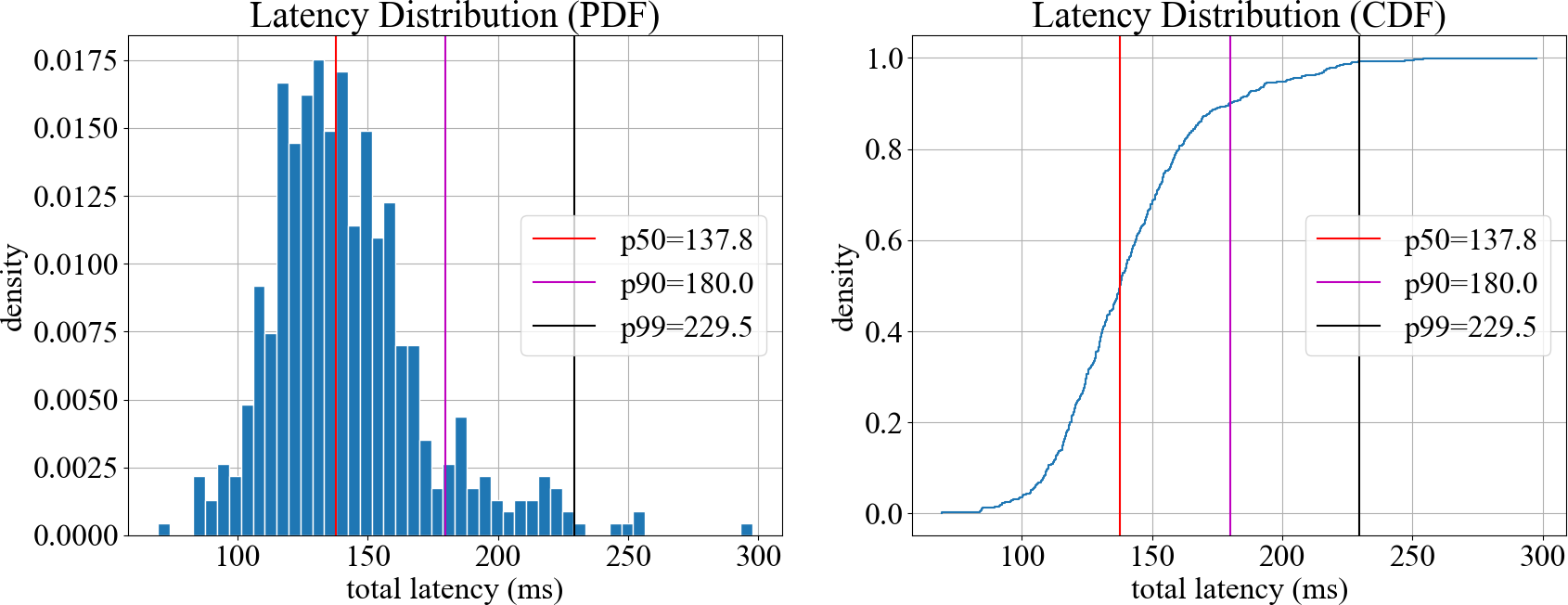}
	\caption{Full Path Access Latency Distribution}
	\label{fig_full}
\end{figure}

\begin{figure}[H]
	\centering\noindent\includegraphics[width=0.48\textwidth]{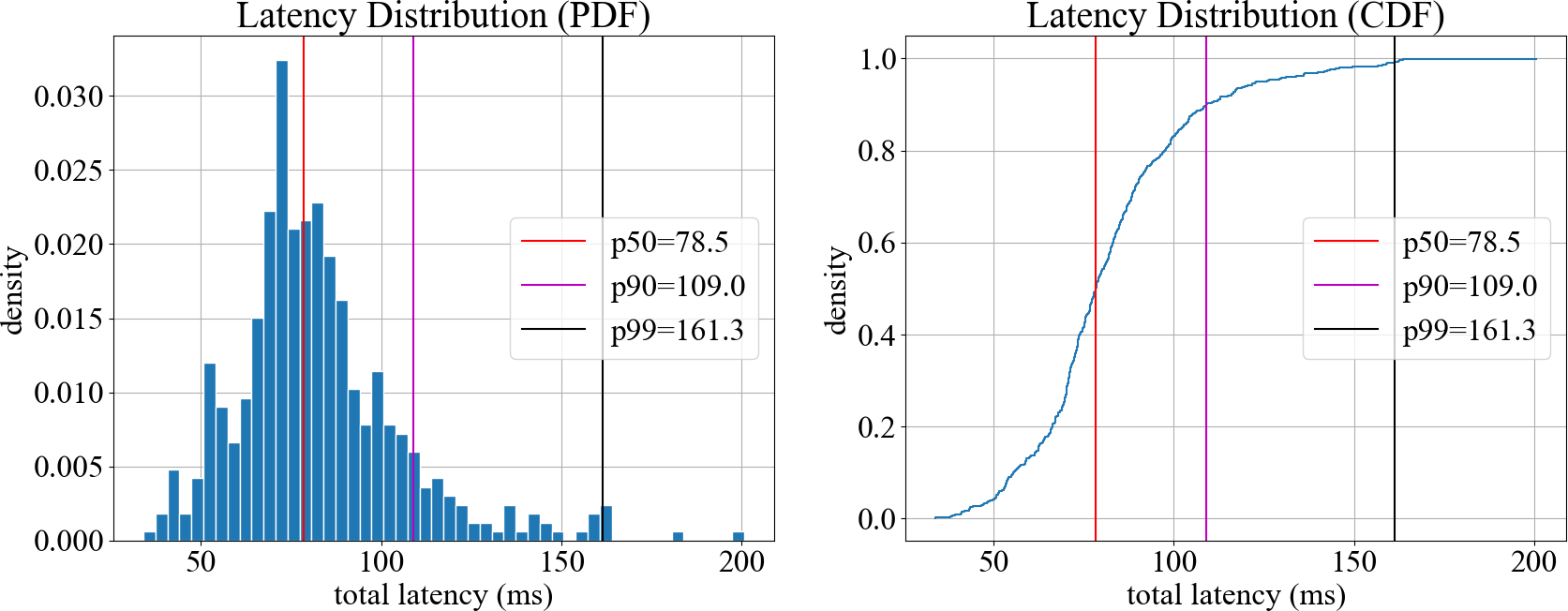}
	\caption{Shortcut Access from Internet Latency Distribution}
	\label{fig_short}
\end{figure}

\begin{figure}[H]
	\centering\noindent\includegraphics[width=0.48\textwidth]{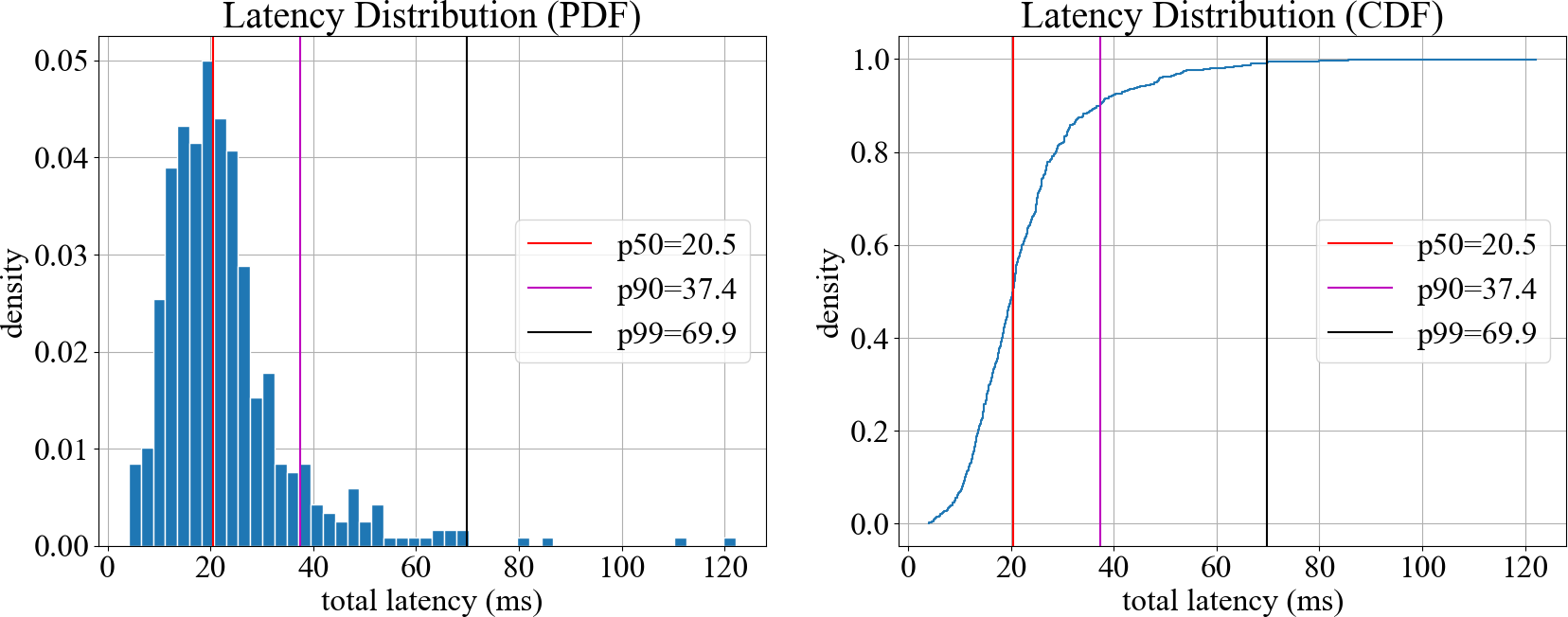}
	\caption{Shortcut Access from Intranet Latency Distribution}
	\label{fig_short2}
\end{figure}

To study the performance of our shortcut strategy, we contrast its performance against the full path approach. Figures \ref{fig_full}, \ref{fig_short} and \ref{fig_short2} represent the latency distributions of 500 times of access validations conducted via full path, shortcut from internet, and shortcut from intranet, respectively. Note that here we do not cold start hub and validators, only client process is restarted every time. 

%The latency distribution of 500 full path accesses:
%The latency distribution of 500 shortcut accesses from internet:
%The latency distribution of 500 shortcut accesses from local network:

The results from the figures show that in the median case ($P_{50}$), utilizing shortcut access over the internet yields a 43\% reduction in access time, while intranet shortcuts offer saving of approximately 73.9\% compared to internet shortcuts. In the worst-case scenario ($P_{99}$), internet shortcut access provides a savings of about 29.7\%, and leveraging intranet shortcuts further saves approximately 56.7\%.

\section{Conclusion}\label{sec:conclusion}
In this paper, we introduce the design of our P2P-BEAC frameowrk, a blockchain-based framework to support multiple access control models in IoT systems. We adapted and extended libp2p to meet the specific needs of NAT-isolated IoT environments and build a virtually universal domain that allows IoT services to be accessed seamlessly by any accessors in the domain. We then outline the  blockchain embedded access control (BEAC) mechanisms in the framework. 

Through the implementation of a bootstrapping process for domains and devices, BEAC ensures efficient recovery and re-synchronization of access control policies after any disruptions, enhancing the system's scalability and resilience.
By integrating BEAC with a range of access control models, our framework proves its flexibility in meeting various operational demands.
Our shortcut mechanism incorporated in BEAC cleverly ensures critical user access remains uninterrupted during internet outages as well as greatly shortens the latency due to access validation.

A thorough performance study for our P2P-BEAC framework is conducted and results show that it achieves significant performance improvements compared to the traditional blockchain based access control schemes. 

\bibliographystyle{IEEEtran}
\bibliography{Thesis}

\end{document}